\documentclass[smallextended , natbib, final]{svjour3}

\usepackage{color}
\usepackage{amsmath}
\usepackage{amssymb}
\usepackage{mathptmx}
\usepackage[pdftex]{graphicx}
\usepackage{subfigure}
\usepackage[misc,geometry]{ifsym}

\journalname{ArXiv}

\begin{document}
 
 \title{The Carbon Cycle as the Main Determinant of Glacial-Interglacial Transitions}
 
 
 \author{Diego Jim\'enez de la Cuesta \and Ren\'e Gardu\~no \and Dar\'io N\'u\~nez \and Beatriz Rumbos \and Carlos Vergara-Cervantes}
 
 
 \institute{
  Diego Jim\'enez de la Cuesta (\Letter) \and Ren\'e Gardu\~no
   \at Centro de Ciencias de la Atm\'osfera, Universidad Nacional
       Aut\'onoma de M\'exico\\
       Circuito Exterior C.U.\\
       M\'exico D.F. 04510\\
       M\'exico.\\
       Tel.: +(52)5556228222 Ext. 44981\\
       \email{diegojdelac@ciencias.unam.mx}
  \and
  Dar\'io N\'u\~nez \and Carlos Vergara-Cervantes
   \at Instituto de Ciencias Nucleares, Universidad Nacional
       Aut\'onoma de M\'exico\\
       Circuito Exterior C.U., PO box 70-543\\
       M\'exico D.F. 04510\\
       M\'exico.
  \and
  Beatriz Rumbos
   \at Instituto Tecnol\'ogico Aut\'onomo de M\'exico\\
       R\'io Hondo No. 1, Col. Progreso Tizap\'an\\
       M\'exico, D.F., 01080\\
       M\'exico.
 }
 
 \date{\today}
 
 \maketitle

 \begin{abstract}
  An intriguing problem in climate science is the existence of Earth's glacial cycles. We show that it is possible to generate these periodic changes in climate by means of the Earth's carbon cycle as the main determinant factor. The carbon exchange between the Ocean, the Continent and the Atmosphere is modeled by means of a tridimensional Lotka-Volterra system and the resulting atmospheric carbon cycle is used as the unique radiative forcing mechanism. It is shown that the carbon dioxide (CO$_{2})$ and temperature anomaly curves, which are thus obtained, have the same first-order structure as the 100 kyr glacial--interglacial cycles depicted by the Vostok ice core data, reproducing the asymmetries of rapid heating--slow cooling, and short interglacial--long glacial ages.
 \end{abstract}
 
 \section{Introduction.}\label{s:intro}
  
  For the last 3 million years the Earth -- and thus its climate -- has transited from having an extensive ice-covered surface, or glacial periods, to intervals with narrow ice cover and milder temperatures, or interglacial ages. Adh\'{e}mar and Croll, both in the XIX century, suggested that Earth's orbital changes could be responsible for the glacial cycles\footnote{Throughout this work {\it cycle} referred to glacial-interglacial cycle, stands for oscillation of Cryosphere extension; on the other hand, {\it cycle} referred to carbon for example, stands for the circulation of carbon in the Earth's system.}, as quoted by \citet{Pail01A}. However, it was not until the following century that M. Milankovi\'{c} took up the study of the orbital theory of climate in a series of seminal papers (\citet{Mila20,Mila30A,Mila41}, as well as \citet{Mila98}).
  
  Milankovi\'{c} recognized the importance of periodic changes in orbital parameters -- mainly eccentricity, obliquity and precession -- in Summer insolation in the Northern Hemisphere. This insolation forcing triggers an ice-albedo \ feedback mechanism, which has been a popular explanation for driving the glacial cycles \citep{HaIS76A,MGLe12A}. Classical examples of ice-albedo models are those by \citet{Budy69A} and \citet{Sell69A}.
  
  The glacial cycles started off as having a 41 kyr period and being essentially symmetric during the Pliocene, but about one million years ago the period changed to 100 kyr, the amplitude became greater and the cycles became assymetric: glaciations develop slowly and warm periods arise in a very short time, geologically speaking. This structural change is referred to as the \emph{mid-Pleistocene transition} \citep{MGLe12A,Huyb07A,AsTz04A,TzGi03A}. Given that oscillation of orbital parameters have remained essentially unchanged, the mid-Pleistocene transition needs to be explained by phenomena other than insolation. In particular, ice sheet dynamics, Ocean circulation, nonlinear responses and carbon dioxide (CO$_{2}$) may also play a strong role in driving glacial cycles \citep{Pail01A,Huyb07A,IkTa99A,Taji98A}. Efforts in order to incorporate this factors into a model have been put forward in classical works by \citet{KaCG79A} and \citet{SaMa88A} or the comprehensive model shown recently by \citet{FoRWA13}.
  
  The role of CO$_2$ as a determinant of climate has been aknowledged since John Tyndall in 1861, who studied the absortion properties of several gases, in particular CO$_{2}$ and water vapor \citep{Herr88}, up to the most recent IPCC (2013) assessment reports. However, its impact over the glacial cycles of the past million years remains somewhat obscure. \citet{Hogg08A} proposed a feedback mechanism in which the carbon cycle is a function of temperature (T). His model manages to reproduce some main features of the glacial cycles; nevertheless, the cycles themselves are still triggered by changes in orbital parameters and carbon is only a feedback on their amplitude. 
  
  In this work we propose a simple model for the carbon cycle that could account for the 100 kyr asymmetric glacial cycles depicted in the data from the Vostok ice core \citep{Peea99A}. The aim is to show that the Earth's ``metabolism'' \citep{Stef00A} could be the main determinant of climate; consequently, CO$_{2}$ variations would be much more than just an amplifying mechanism in the climate system, but rather the main driving factor. In this way, this work contributes in underlying the importance of CO$_2$ in producing climate changes \citep{Shea13A}.
  
  Following the work of the International Geosphere and Biosphere Programme, we may think of the Earth's Biogeophysicochemical processes in a holistic way \citep{Faea00A,Stef00A}. The ice core records, such as Vostok's \citep{Peea99A}, depict $T$, CO$_{2}$ and methane data that resemble a rythmic metabolic pattern. Just as the Earth's short term orbital/insolation features are embedded into our human metabolism, the long term orbital/insolation changes must be embedded into the Earth's ``metabolism''. It is the planet's ``metabolic'' process that is proposed as the primary driving factor for the climate system We must emphasize that, when we use the word ``metabolism'' in relation to the Earth, we are doing an analogy and Earth's ``metabolism'' is not only biological in nature, but chemical and physical too. \citet{Stef00A} presents a simple qualitative description about how this complex system works. We shall bypass the complexities to construct a simple model describing the CO$_{2}$ and $T$ oscillations from the time series extracted from the Vostok ice cores.
  
 \section{The Vostok ice core time series}\label{s:VICTS}
  
  Here we give a brief description of the Vostok ice core time series, we shall concentrate exclusively on the CO$_{2}$ and temperature anomaly ($\Delta T$) series.
  
\begin{figure*}
   \centering
   \includegraphics*[viewport=0mm 0mm 161mm 80mm, width=1.0\textwidth]{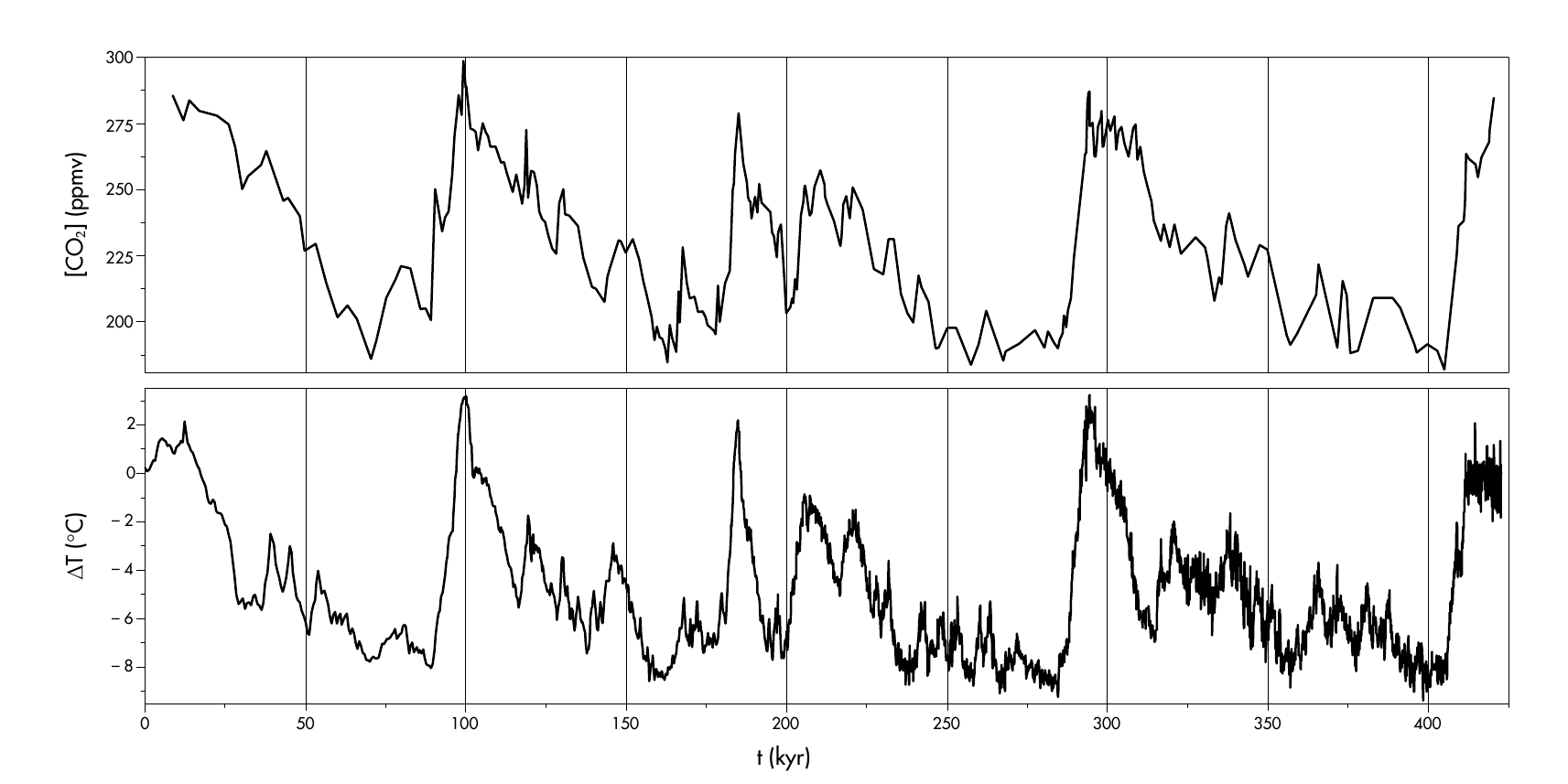}
   \caption{Two time series from the Vostok data, \citet{Peea99A}. $\Delta T$ and concentration of CO$_{2}$. The right hand side corresponds to modern times and the left hand side to the core bottom or oldest data. Time scale is in kiloyears (kyr).}
   \label{fig1}
  \end{figure*}

  There are two time series belonging to the variables of interest: CO$_{2}$ in ppmv and $\Delta T$ (with respect Antarctic Vostok station recent climatic normal temperature), in degrees Celsius, extracted using $\delta \,^{18}\mathrm{O}$ and $\delta \,\mathrm{D}$ as proxies as described in \citet{Peea99A}. In figure~\ref{fig1} we present both time series with time labeled from $0$ (the deepest data and therefore the oldest one) to the most recent data. Data and the depth-age correlation were downloaded from NOAA-NGDC \citep{Peea99A}.
  The time series display the following well known striking features \citep{MGLe12A,Peea99A,Stef00A,Faea00A}:
  
  \begin{enumerate}
   \item Both are periodic: There are four cycles with five maxima and four minima regularly distributed along temporal record.
   \item Both are (almost) in phase \citep{GaCrT05A}, that is, main and secondary extrema are matched over time. This implies a correlation between $\Delta T$ (and therefore $T$) with the concentration of CO$_{2}$ along the milenia time scale.
   \item Both are bounded. This feature, together with periodicity, shows that the climatic system oscillates between two extremal states along the time record analysed \citep{Peea99A,Stef00A}.
   \item There is an assymetry in the cycles: Heating seems to occur very fast while cooling is slow.
   \item Closely related with the previous assymetry, we observe that interglacials are short but glacial periods are long.
  \end{enumerate}
  
  A qualitative explanation of the glacial-interglacial cycles \citep{Faea00A,Stef00A} may be described in a short stylized version as follows: starting near a peak in the time series, surface $T$ and atmospheric CO$_{2}$ are at their highest. Increased precipitation causes a surge in Continental biomass, then carbon is transferred to Continent (emerged lands) from Atmosphere (troposphere) until a saturation level is reached. At that point, triggered by runoff, the Ocean (saline waters that surround Continent) absorbs CO$_{2}$ while it gradually recovers biomass. $T$ and CO$_{2}$ drop until the Ocean reaches its saturation level and the process is now reversed. In this scenario, this ``control switching'' between Ocean and Continent -- mediated by Atmosphere, like a counduit -- is the main driver of the glacial cycles.
  
 \section{Earth system model}\label{s:model}
  
  \subsection{Modelling the carbon cycle.}\label{ss:carbon}
   
   The idea behind our carbon cycle model is simplicity. In a striking opposition to Atmosphere, Continental and Oceanic carbon reservoirs are driven by very rich and complex internal relationships due to existence of biological, chemical and physical processes, which transform their carbon contents between several chemical compounds. Thus, each reservoir can be thought globally as a great box with subreservoirs. Since we want a model without intrincate details, our three carbon stocks shall be that of those subreservoirs where carbon is available for exchange among the three spaces: Ocean, Continent and Atmosphere.
   
   For such choice, our model is not a chemical one in the following sense: system does not need to obey conservation of mass at all because Oceanic and Continental stocks are not the whole quantities of carbon that exist in Ocean and Continent, therefore the system from our perspective is open.
   
   Now, relationships between reservoirs can be of three sorts: those controlled by donor stock, those controlled by receptor stock and those controlled by stocks of both (a true interaction).
   
   Since this model is simple, phenomena driving fluxes will not be described in detail. That is, we will not model each single mechanism, which drives a particular flux, but only a schematic joint representation. Phenomena, therefore, will be described by constants associated to each relationship we have referred formerly. Since some processes could suffer changes in sign dependent on time -- as it was told in section \ref{s:VICTS} -- there will be a corrective term that is mathematical in nature. Once we have outlined the principles for constructing model, we proceed.
   
   Continental stock, which is represented by the symbol $C_{1}$, relates with Atmospheric one (represented by $C_{2}$) through processes that rely on biological activity -- dependent on the carbon available for it in $C_{2}$ -- which are physical and chemical in essence, such as photosyntesis and respiration. In this way, there is some term with the product $C_{1}C_{2}$.
   
   Similarly, relationship of Oceanic (which we represent as $C_{3}$) and Atmospheric stocks must have the same form, here the main processes being dissolution and outgassing. Then there exists a term with the product of $C_{2}C_{3}$.
   
   For Continental and Oceanic stocks, the relationship is not so straightforward. First, Continental carbon goes to Oceanic stock through runoff: so there will be a term with $C_{1}$. On the other hand, not all the carbon transferred to Ocean in that way is readily available for exchange. Therefore, it does not enter directly to $C_{3}$ reservoir: this carbon can be used by Oceanic biosphere or take part into acidity-alcalinity balance, among interactions of other nature. Thus, carbon input of Continental origin is, at the end, effectively controlled by Ocean itself when it comes to talking about our stock $C_{3}$. Then there exists a term $C_{3}$. 
   
   Once we have described the possible relationships among our carbon reservoirs, we can see that interaction between Ocean, as well as Continent, with Atmosphere could have sign changes (which represent a reversal of overall process), as told previously in section \ref{s:VICTS}. Since these interactions involve Atmosphere, we need to introduce a corrective term, which is only $C_{2}$-dependent, and shall account for some excess or defect of carbon due to simplification.
   
   Thus, temporal change in $C_{1}$ will be given by runoff loss into the Ocean and interaction with $C_{2}$, while in $C_{3}$ it will be determined by incorporation of runoff carbon from $C_{1}$, controlled by $C_{3}$ itself -- as we referred -- plus interaction with $C_{2}$. Finally, $C_{2}$ temporal change is described by interactions with both $C_{1}$ and $C_{3}$ plus the corrective term. Therefore, the following ODE coupled system results
   \begin{align}
    \begin{cases}
     \dot{C}_{1}=-\alpha C_{1}+\beta C_{1}C_{2}, \\ 
     \dot{C}_{2}=\gamma C_{2}-\beta C_{1}C_{2}+\varepsilon C_{2}C_{3}, \\ 
     \dot{C}_{3}=\eta C_{3}-\varepsilon C_{2}C_{3}.
    \end{cases}
    \label{CC}
   \end{align}
   where $\alpha,\beta,\gamma,\varepsilon,\eta>0$ are our simplified representations of the processes that establish carbon fluxes and are constants as we stated before. Prescribed signs to each term are negative when there is a carbon cession. Interaction terms have an skew-symmetric look; since any term that appears in one equation, also appears in another one with reversed sign.
   
   The model -- as was constructed -- allows for extensions. Any extension would rely on additional constitutive equations that link coefficients of ODE system directly with physical, chemical and biological principles behind fluxes.
   
   It could be verified that this ODE system have a particular structure: $n-$dimen\-sional Lotka-Volterra system. Lotka-Volterra systems have arisen first in fields as diverse as interaction of biological populations (due to Vito Volterra) and chemical autocatalytic reactions (by Alfred J. Lotka).
   
   The solutions of a Lotka-Volterra system could be rich. However, we expect periodic solutions as we have seen in previous section \ref{s:VICTS}. For our particular dynamical system, the condition for periodic solutions is that the ``survival'' ratio for $C_{1}$, $-\alpha /\beta$, equals the ``survival'' ratio for $C_{3}$, $\eta /(-\varepsilon) $. This implies $\eta =\alpha \,\varepsilon /\beta $ in order to assure periodic solutions \citep{CPPW02A}.
   
   One must note that interaction matrix for Lotka-Volterra system in this case (that formed with the coefficients of the terms with products) is skew-symmetric. It can be proved that a system with this structure admits conserved quantities, or ``constants of motion'' in mechanics. Moreover, these systems have a Hamiltonian structure \citep{Plan95A}.
   
   Conserved quantities depend on the fixed points of the system, and in our case we can see that there is a one-parameter family of constants of motion since there is a line of fixed points. Hamiltonians could be obtained by introducing Volterra's conserved quantity, as \emph{ansatz} or educated guess. Then we solve for its time derivative to vanish identically, as we show at the appendices. One member of the family of conserved quantities is
   \begin{align}
    H(t)=C_{1}+C_{2}+C_{3}-\ln\left(C_{1}^{\frac{\alpha}{\beta}}C_{2}^{\frac{\gamma}{\beta}}\right).
    \label{Hamiltonian}
   \end{align}
   See \ref{s:LVH} for details.
   
  \subsection{Modelling the energy chain.}\label{ss:energy}
   
   By energy chain we mean an energy balance. Jointly, energy chain and carbon cycle make up our Earth System model. Although we again seek simplicity, abstraction is lesser than in carbon cycle model.
   
   For the sake of simplicity, we assume a constant radiative input from the Sun. Also, the Earth is thought as a spherical blackbody for longwave radiation with the Atmosphere being ``gray'' to it.
   
   Incoming solar radiation reaches Earth, part of which is reflected towards space due to planetary albedo. Radiation, which was not reflected, is absorbed and then reemitted as longwave radiation. For modelling purposes, we consider two reservoir variables: $S$ for Surface (Ocean and Continent together) energy density and $A$ for Atmosphere energy density (both in $\mathrm{Jm}^{-2}$. Atmospheric one is integrated over the vertical column).
   
   We use the symbol $\Omega_{\odot}$ for the solar incoming flux (in $\mathrm{Wm}^{-2}$) -- which is one fourth of total solar irradiance $I_{\odot}$ -- and $a_{1}$ for planetary albedo, $a_{2}$ for the fraction of non-radiative energy from Surface that drives thermal processes, $a_{3}$ for the fraction of Surface radiative flux absorbed by Atmosphere (the rest is outgoing radiation towards space) and $a_{4}$ for the fraction of Atmospheric radiative flux that is absorbed by Surface (again, the rest goes to space). These are all the processes we take into account.
   
   Similarly as we have done with carbon cycle, temporal change on $S$ equals the solar input minus albedo fraction, plus the fraction $a_{4}$ of Atmospheric radiative flux and, since all energy absorbed by Surface is reemitted, minus the Surface radiative flux. By analogy, temporal change on $A$ equals the fraction $a_{2}$ of energy from Surface plus the fraction $a_{3}$ of Surface radiative flux (complemetary fraction $(1-a_{2})$) and minus the Atmospheric radiative flux.
   
   Surface and Atmospheric stocks -- $S$ and $A$ -- are proportional to fluxes, since it would be unrealistic to say that there is some sort of energy acummulation in our reservoirs. That means, every fraction has a factor with time dimensions: $1[t]$, making the resulting balance ODE system unit-consistent. Moreover, we can take $\Omega_{\odot}$ as a nondimensionalization factor giving us the following energy chain model
   \begin{align}
    \begin{cases}
     \dot{S}=(1-a_{1})+a_{4}\,A-S, \\ 
     \dot{A}=(a_{2}+a_{3}\,(1-a_{2}))S-A,
    \end{cases}
    \label{EC}
   \end{align}
   This system is linear, but coefficients shall not be constant. Fractions like planetary albedo, need to be variable for having something meaningful over this time scales. How vary this fractions?
   
   For example, albedo depends on the extension of Cryosphere, together with other factors that are strongly associated such as cloud cover, vegetation cover and desertification. An indicator of Cryosphere presence or absence is $T$, however this variable is what we want to model with energy chain. We can think that Atmospheric carbon is another good indicator of Cryosphere extension, since greenhouse gases (GHG) play the main role in modifying radiative transfers. We use results showing that global mean albedo decreases in response to CO$_{2}$ feedback \citep{Bend11A}. Similarly, one can reason for the remaining parameters, in particular this relationship is stronger for $a_{3}$ and $a_{4}$.
   
   Now, we can ask for constitutive equations relating parameters with Atmospheric carbon. Saturation, smoothness and boundedness of these processes are fundamental for seeking a meaningful -- yet simple -- constitutive law. For example, GHG radiative absortion is bounded -- since the increment of their concentration does not mean a proportional increment in radiative forcing --. In other words, there is a saturation limit where adding a lot of GHG results in a tiny increment of their radiative effect. One such functions that have this characteristics is logistic function, which is logistic ODE solution. Thus, we shall assume that $a_{i}$ are dependent on $C_{2}$ through the following constitutive relations
   \begin{align}
    a_{i}(C_{2})=\frac{1}{1+\left(\frac{1}{a_{i0}}-1\right)e^{-r_{i}\,(C_{2}-C_{2_{\mathrm{ref}}})}},
    \label{par}
   \end{align}
   where $r_{1}<0$ (reverse feedback) and $r_{2,3,4}>0.$ $C_{2_{\mathrm{ref}}}$ is a known quantity of CO$_{2}$, and we use a recent value in order to determine a particular solution for the coefficients $a_{i}$.
   
   From this we are ready to calculate Surface mean $T$ (in $^{\circ}\mathrm{C}$) using Stefan--Boltz\-mann law as:
   \begin{align}
    T(t)=\left( \frac{(1-a_{2})\,S\,\Omega_{\odot }}{\sigma }\right) ^{\frac{1}{4}}\,-\,273.15
    \label{temp}
   \end{align}
   where $\sigma $ is Stefan--Boltzmann constant, and $S$ is the nondimensionalized $S$.
   
   These somewhat unusual equations, compared with those in related literature, could be transformed to a more standard energy balance equation as we show in \ref{s:ECC}.
   
 \section{Simulation and results}\label{s:sim}
  
  The Earth system model given by equations~(\ref{CC}) and (\ref{EC}), together with~(\ref{par}), was solved numerically. We searched for reasonable initial conditions, probed parameter values (alone and grouped) for finding their effects over system's evolution and, after several iterations, we found the best fitting parameters. For this task we used mean square error between observations and simulations as a guide. For the formulas linking model and observation spaces, see \ref{s:metsum}.
  
  The initial conditions for the energy chain ($S_0$ and $A_0$) were extracted and modified from similar problems in the literature \citep{DeWi00}. Additionally, considering the recent value of $C_{2_{\mathrm{ref}}}=387\,\mathrm{ppmv}$, we also fix $a_{i,0}$: $a_{1,0}=0.313,a_{2,0}=0.207,a_{3,0}=0.897,a_{4,0}=0.624$.
  
  Unfortunately, initial conditions for the carbon cycle were difficult to establish although there are proxies for Continental and Oceanic stocks, they are not in the form we need. Given our hypotheses -- Atmosphere acts as a conduit between Continental and Oceanic carbon, our carbon stocks are only those quantities available for exchange and the fact that we begin at an interglacial period with an active Continent and a dormant Ocean -- we may infer the following ordering for the initial values of the carbon stocks: $C_{3,0}\leq C_{1,0}\leq C_{2,0}$.
  
  Results are presented in table \ref{tp}.
  \begin{table}
   \centering
   \resizebox{8.5cm}{!}{
   \begin{tabular}{c|c||c|c}
    \hline
    Initial conditions & Value & Parameters & Value\\
    \hline
    $C_{1,0}$ &0.2305 &$\alpha$ &0.02928\\
    $C_{2,0}$ &0.6295 &$\beta$ &0.1050\\
    $A_{0}$ &1.260&$\varepsilon$ &0.3630\\
    $S_{0}$ &1.388&$r_{1}$ &$-4.122\,\times 10^{-3}$\\
    - &- &$r_{2}$ &$4.135\,\times 10^{-3}$\\
    - &- &$r_{3}$ &$4.76\,\times 10^{-4}$\\
    - &- &$r_{4}$ &$7.4\,\times 10^{-5}$\\
    \hline
   \end{tabular}
   }
   \caption{Initial conditions and best fitting parameters for carbon cycle and energy chain.}
   \label{tp}
  \end{table}
  Arbitrarily, initial conditions fulfill $C_{1,0}+C_{2,0}+C_{3,0}=1$. We also found that $\gamma$ must be
  \begin{align}
   \gamma=\alpha\frac{C_{1,0}}{C_{2,0}}-\frac{\alpha\varepsilon}{\beta}\frac{C_{3,0}}{C_{2,0}}.
  \end{align}
  
\begin{figure*}
   \centering
   \includegraphics*[viewport=0mm 0mm 159mm 44mm, width=1.0\textwidth]{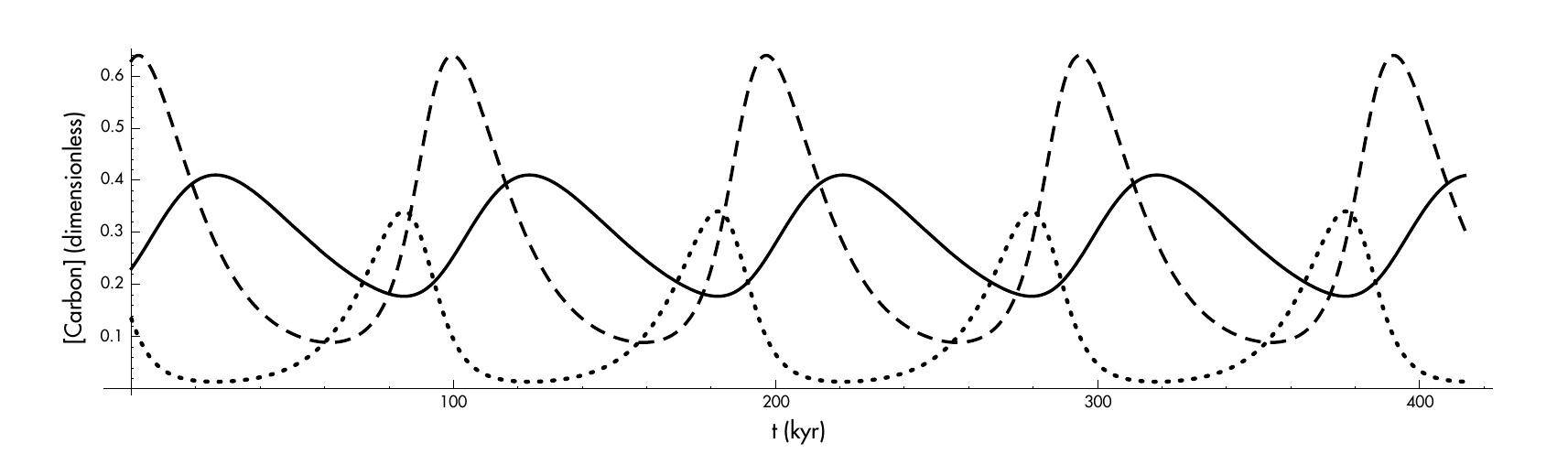}
   \caption{Carbon stocks from simulation. Solid $C_{1}$, dashed $C_{2}$ and dotted line $C_{3}$.}
   \label{fig2}
  \end{figure*}

  Computed time series for the three carbon stocks are depicted in figure~\ref{fig2}. We observe that the phase shift between $C_{1}$ and $C_{3}$ mimics the transition between Continental and Oceanic control: one of these carbon stocks becomes almost depleted when the other one is dominant. This does not happen with atmospheric carbon $C_{2}$ which is always, figuratively speaking, alive. 
  
\begin{figure*}
   \centering
   \includegraphics*[viewport=0mm 0mm 165mm 43mm, width=1.0\textwidth]{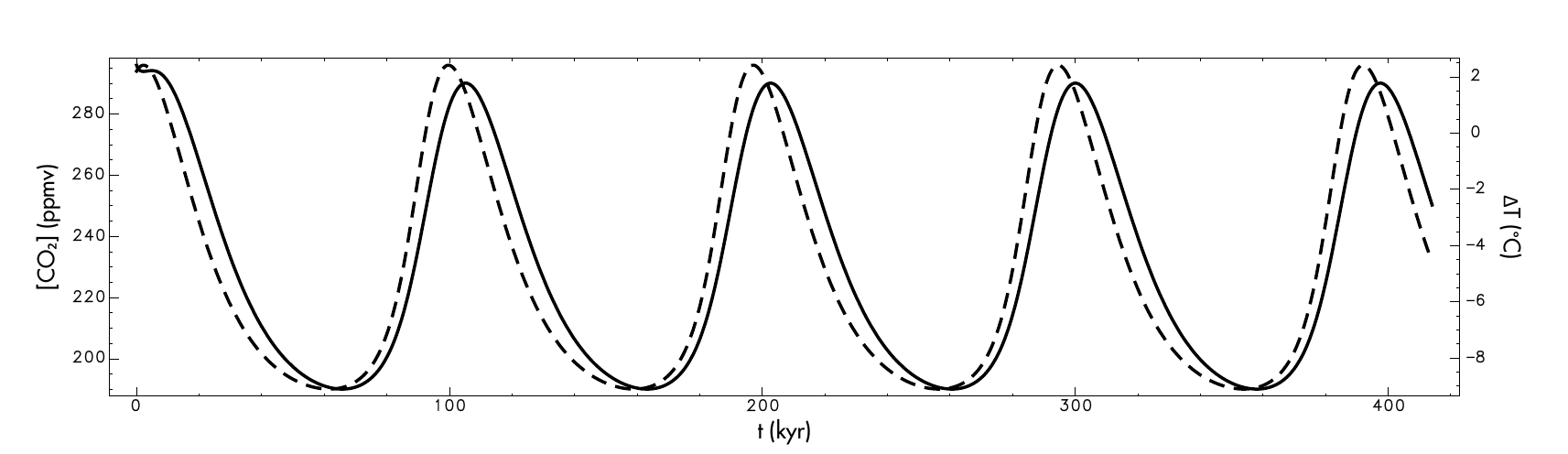}
   \caption{Simulated Atmospheric carbon (dashed line) and $\Delta T$ (solid line). Ordinate axis for carbon is concentration of carbon dioxide in parts-per-million by volume and $\Delta T$ in $^{\circ}\mathrm{C}$.}
   \label{fig3}
  \end{figure*}

  We must emphasize that the purpose of this model is not to be an accurate fit for the Vostok time series, but rather to reproduce first-order features: periodicity and assymetries. The computed time series for Atmospheric CO$_{2}$ and $\Delta T$ depicted in figure~\ref{fig3} show that this is indeed the case, with faster heating and slow cooling, and interglacials of shorter duration than glacial ones.
  
\begin{figure}
   \centering
   \includegraphics*[viewport=4mm 2mm 162mm 95mm, width=1.0\textwidth]{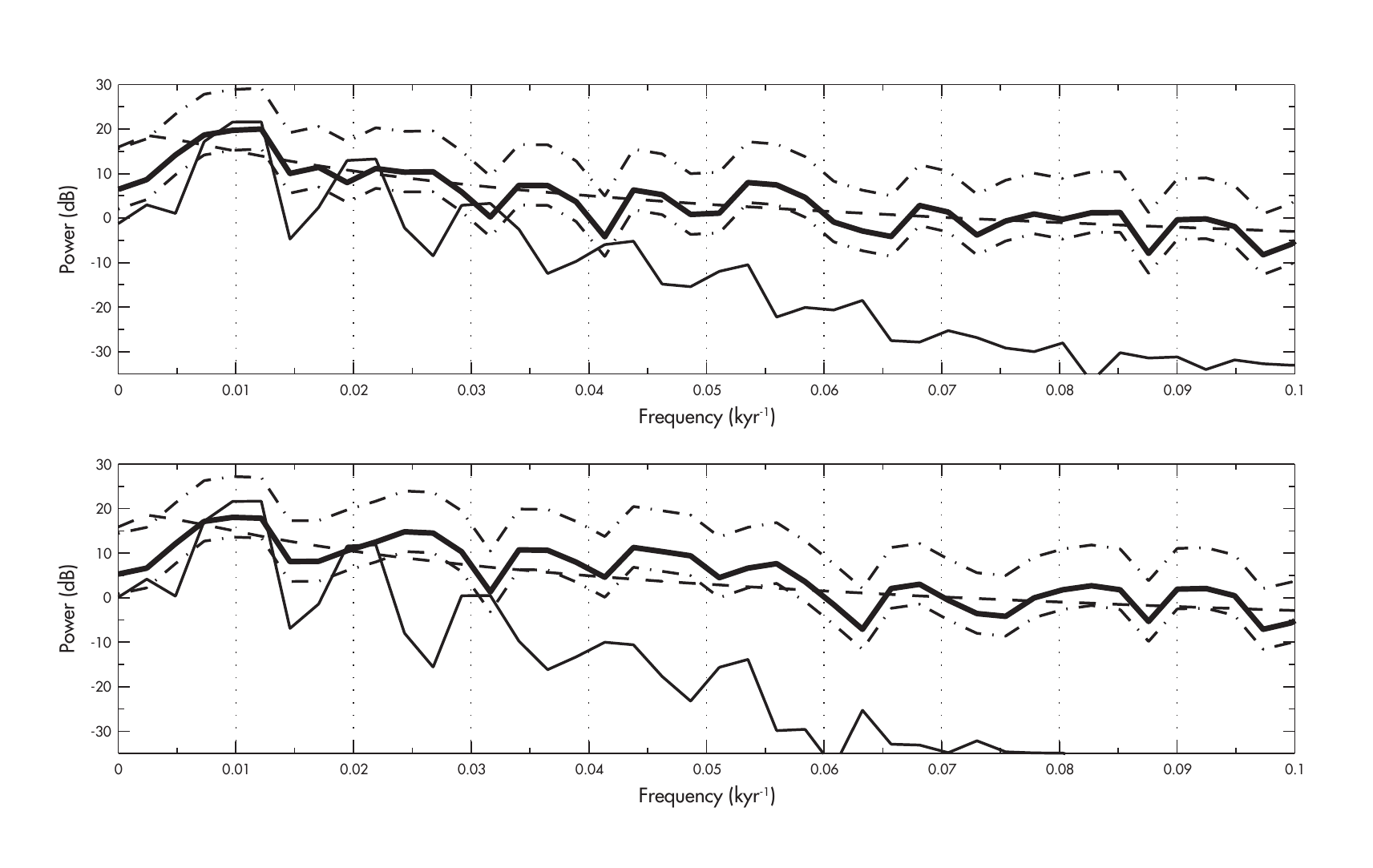}
   \caption{Periodograms for carbon (top) and $\Delta T$ (bottom). Heavy continuous line is observed, light continuous is simulation. Dash-dotted lines are the 0.95 confidence interval and dashed line is noise baseline, both for observed data.}
   \label{fig4}
  \end{figure}

  On analyzing the (frequency) power spectrum of the simulated series (see \ref{s:metsum} for details on power spectra methods), several points stand out. The CO$_{2}$ series presents a dominant frequency around 105 kyr, followed by the 46 kyr and 21 kyr frequencies. This closely agrees with observed data \citep{MGLe12A}, which is not surprising as the model parameters were calibrated with the Vostok data. Nevertheless, it reassures that this simple model does indeed capture the main frequencies from orbital forcing of the climate system in the time period covered by the data. The power spectrum of the simulated $\Delta T$ has the same dominant frequencies, a striking fact, considering that the carbon cycle is the only driving mechanism for the $T$ changes. Thus, the principal frequencies are obtained without any reference to the Earth's orbital parameters. Figure~\ref{fig4} depicts these considerations.
  We do not mean that orbital parameters are irrelevant, but rather that they are already incorporated into the Earth's carbon metabolism which is the driver of our model. For example, life is forced by these changes in orbital parameters and biology is fundamental for a carbon cycle to exist.
 \section{Conclusions}\label{s:conc}
  
  The model depicted here differs from existing glacial cycle models in one crucial aspect: it does not consider orbital parameters at all; thus, changes in insolation, either global or local, are not explicitly taken into account, as we hypothesise that their main impact on climate is through their effect within the carbon cycle. Our model shows that it is possible for the carbon cycle to drive glaciations and deglaciations. In this scenario, the behavioural change of the mid-Pleistocene transition may be explained by a change in carbon cycle fluxes, which would bring about a change in one or more of the model parameters. In fact, this is probably the case in the lapse covered by the Vostok data, as one can observe anomalies like the last two interglacials: one of them is reached before than expected and the last one comes later than expected.
  
  We show that it is possible for the glacial cycles to be driven by long term exchanges of carbon between Oceanic and Continental reservoirs, with the Atmosphere mediating the process. By introducing Atmospheric carbon content as a forcing factor in the energy chain, we obtain the desired cyclic $\Delta T$ pattern typical of the Vostok time series. Thus, the main conclusion is that glacial cycles could be essentially carbon driven and insolation triggered.
  
  The carbon cycle is implicit in the Oceanic-Continental control-switching and our model attempts to make a first-order simulation of long term cycles (100 kyr in the Vostok time series case) within the carbon cycle and energy chain. A consequence of our model is that these long term CO$_{2}$ cycles should have the same periodicity for all three carbon stocks: Atmospheric, Continental and Oceanic.
Unfortunately, we would need to devise a method to compare our carbon stocks with quantities found in past and present proxies for Continental and Oceanic carbon, comparison not yet done by us.
  
  Here we pictured Earth's metabolism as the long term carbon exchange between Oceanic and Continental reservoirs. The Atmosphere served only as an intermediary in this process; nonetheless, one can not help but worry about the possible effect of the present value concentration of greenhouse gases in the Atmosphere. Altering the Earth's complex metabolic process is not something we should want to experiment with.
  
%
  
 \appendix
 \section{Conservation quantity in Lotka-Volterra system}\label{s:LVH}
  
  We know that Volterra's constant of motion has the form
  \begin{align*}
   H=C_{1}+C_{2}+C_{3}-\ln\left(C_{1}^{A}C_{2}^{B}C_{3}^{\Gamma}\right),
  \end{align*}
  where $A,B,\Gamma$ are related with system equilibria. In this case equilibria lie over a line and then are a one-parameter family, so one of the three exponents is free. Since $H$ is conserved, its temporal derivative must vanish identically.
  \begin{align*}
   \dot{H} 
    &=\dot{C}_{1}+\dot{C}_{2}+\dot{C}_{3}-\left(A\frac{\dot{C}_{1}}{C_{1}}+B\frac{\dot{C}_{2}}{C_{2}}+\Gamma\frac{\dot{C}_{3}}{C_{3}}\right)
  \end{align*}
  
  Thus, if we insert equations given by system (\ref{CC}) in the derivative of $H$ we obtain
  \begin{align*}
   \dot{H} &=-\alpha C_{1}+\beta C_{1}C_{2}+\gamma C_{2}-\beta C_{1}C_{2}\\
    &+\varepsilon C_{2}C_{3}+\frac{\alpha\varepsilon}{\beta}C_{3}-\varepsilon C_{2}C_{3}\\
    &+\alpha A-\beta AC_{2}-\gamma B+\beta B C_{1}\\
    &-\varepsilon BC_{3}-\frac{\alpha\varepsilon}{\beta}\Gamma+\varepsilon\Gamma C_{2}\\
    &=\left(\alpha A-\gamma B-\frac{\alpha\varepsilon}{\beta}\Gamma\right)+\left(-\alpha+\beta B\right)C_{1}\\
    &+\left(\gamma-\beta A+\varepsilon\Gamma\right)C_{2}+\left(\frac{\alpha\varepsilon}{\beta}-\varepsilon B\right)C_{3}
  \end{align*}
  then, as $H$ temporal derivative vanishes identically, each coefficient must vanish unless we wish that they were the trivial solutions $C_{i}\equiv 0$, which is not the case. Therefore, we have a linear system given by
  \begin{align*}
   \begin{cases}
    \alpha A-\gamma B-\frac{\alpha\varepsilon}{\beta}\Gamma &=0\\
    -\alpha+\beta B &=0\\
    \gamma-\beta A+\varepsilon \Gamma &=0\\
    \frac{\alpha\varepsilon}{\beta}-\varepsilon B &=0
   \end{cases}
  \end{align*}
  Second and fourth equations are one and the same, and yield $B=\frac{\alpha}{\beta}$. From third one we can solve for $A$ or $\Gamma$. Doing this for $A$, as an example, yields
  \begin{align*}
   A &=\frac{\gamma+\varepsilon\Gamma}{\beta}
  \end{align*}
  and substitution of $A$ and $B$ into the first equation of linear system results in
  \begin{align*}
   \alpha A-\gamma B-\frac{\alpha\varepsilon}{\beta}\Gamma &=0\\
   \frac{\alpha\gamma+\alpha\varepsilon \Gamma-\alpha\gamma-\alpha\varepsilon\Gamma}{\beta} &=0
  \end{align*}
  but this vanishes with any value of $\Gamma$. The same happens if we solve for $\Gamma$ in the third one. This is the effect of having a line of fixed points.
  
  Finally the Hamiltonian, or the family of Hamiltonians, is given by
  \begin{align*}
   H(t)=C_{1}+C_{2}+C_{3}-\ln\left(C_{1}^{\frac{\gamma+\varepsilon\Gamma}{\beta}}C_{2}^{\frac{\alpha}{\beta}}C_{3}^{\Gamma}\right)
  \end{align*}
  For $\Gamma=0$ this yields
  \begin{align*}
   H(t)=C_{1}+C_{2}+C_{3}-\ln\left(C_{1}^{\frac{\gamma}{\beta}}C_{2}^{\frac{\alpha}{\beta}}\right)
  \end{align*}
  which is equation (\ref{Hamiltonian}).
  
  For more details see \citet{Plan95A}. Theorem 4.3 in this work gives Volterra's constant of motion and the associated conditions, which the system must satisfy, for such constant of motion to exist.
  
  One can read the first section from third chapter of \citet{Baig10} -- which are course notes on Lotka-Volterra dynamics -- about coordinate transformation which proves that this type of Lotka-Volterra equations have Hamiltonian structure. Volterra named these new coordinates \emph{Quantity of life}.
  
  It is useful to give an explanation on the form of $H$ in the context of our model. Since logarithm could be positive or negative and in our model mass is not conserved -- since we are taking an open system -- then the logarithmic term could be interpreted as the excess (or deficit) of carbon for our system respect the closed system, where total carbon is a conserved quantity.
  
 \section{Energy Balance equations.}\label{s:ECC}
  
  Equation for $S$ with dimensions reads
  \begin{align*}
   \dot{S}=(1-a_{1})\Omega_{\odot}+a_{4}A-S
  \end{align*}
  
  Since $a_{2}(C_{2})$ is a solution of logistic equation, we can write down
  \begin{align*}
   a'_{2} &=r_{2}a_{2}(1-a_{2})\\
   \therefore \dot{a}_{2} &=r_{2}a_{2}(1-a_{2})\dot{C}_{2},
  \end{align*}
  where we have used chain rule.
  
  On the other hand, Surface absolute $T$, is
  \begin{align*}
   T=\left(\frac{1-a_{2}}{\sigma}S\right)^{\frac{1}{4}}
  \end{align*}
  or
  \begin{align*}
   \sigma T^{4}=(1-a_{2})S.
  \end{align*}
  Thus, differentiating both sides with respect to time
  \begin{align*}
   4\sigma T^{3}\dot{T}=(1-a_{2})\dot{S}-\dot{a}_{2}S
  \end{align*}
  and substituting $\dot{S}$ and $\dot{a}_{2}$
  \begin{align*}
   4\sigma T^{3}\dot{T} &=(1-a_{2})(1-a_{1})\Omega_{\odot}+(1-a_{2})a_{4}A\\
    &-(1-a_{2})S-r_{2}a_{2}(1-a_{2})\dot{C}_{2}S\\
    &=(1-a_{2})[(1-a_{1})\Omega_{\odot}+a_{4}A-S\\
    &-r_{2}a_{2}\dot{C}_{2}S]\\
    &=(1-a_{2})[(1-a_{1})\Omega_{\odot}+a_{4}A\\
    &-(1+r_{2}a_{2}\dot{C}_{2})S],
  \end{align*}
  then dividing by $(1-a_{2})$
  \begin{align*}
   \frac{4\sigma T^{3}}{1-a_{2}}\dot{T}=(1-a_{1})\Omega_{\odot}+a_{4}A-(1+r_{2}a_{2}\dot{C}_{2})S.
  \end{align*}
  But $S=\sigma(1-a_{2})^{-1}T^{4}$; thus,
  \begin{align*}
   \frac{4\sigma T^{3}}{1-a_{2}}\dot{T}=(1-a_{1})\Omega_{\odot}+a_{4}A-\frac{1+r_{2}a_{2}\dot{C}_{2}}{1-a_{2}}\sigma T^{4}
  \end{align*}
  or equivalently
  \begin{align}
   4\frac{S}{T}\dot{T} &=(1-a_{1})\Omega_{\odot}+\left(a_{4}A-\frac{r_{2}a_{2}\dot{C}_{2}}{1-a_{2}}\sigma T^{4}\right)\nonumber\\
    &-\frac{1}{1-a_{2}}\sigma T^{4}
  \end{align}
  
  Standard energy balance equation reads \citep{Hogg08A}
  \begin{align}
   R\dot{T}=S_{std}+G_{std}-\sigma T^{4}
  \end{align}
  where $R$ is surface heat capacity, $S_{std}$ is insolation, $G_{std}$ is greenhouse effect contribution.
  
  By identifying terms and comparing, we find that
  \begin{align}
   R &=4\frac{S}{T}\\
   S_{std} &=(1-a_{1})\Omega_{\odot}\\
   G_{std} &=a_{4}A-\frac{r_{2}a_{2}\dot{C}_{2}}{1-a_{2}}\sigma T^{4}
  \end{align}
  and the last term of $G_{std}$ might be interpreted as our $T$ takes into account that some energy from $S$ drives thermal processes. Our Greenhouse effect term is depending on radiative forcing of atmospheric GHG and its evolution. Therefore, we do not prescribe $G_{std}$, or at least it has a strong binding with our carbon cycle. Of course, evolution for $A$ is needed since Atmospheric radiation absorbed by Surface completes the notion of greenhouse effect.
  
 \section{Methods Summary.}\label{s:metsum}
  
  The observed carbon and $\Delta T$ data need to be transformed to fit our model space. In the case of carbon, we take into account that parameters involved are coupled; hence, it is not possible to obtain a specific amplitude given a certain periodicity. With this in mind, we performed a series of iterations in order to obtain a periodicity similar to the observed one. We then used a translation and a homothetic transformation on the observed data to obtain an initial approximation for the amplitude and zero level in the model space.
  
  The results from this iterative process yield the following expression:
  \begin{align}
   C_{2,model}=5.202\times\,10^{-3}\,(C_{2,obs})-0.9
  \end{align}
  
  Similarly, for $\Delta T$, we assumed an annual global climatic normal of 15$\,^{\circ}\mathrm{C}$, that is,
  \begin{align}
   T_{model,celsius}=\delta T_{obs}+15\,^{\circ}\mathrm{C}
  \end{align}
  
  All iterations and simulations were carried out with \textit{Scipy} libraries, and \textit{Mathematica} was used for final numerical integration and graphical output.
  
  Spectral analysis was done with \textit{Matlab} spectral estimation tools. Observed and simulated data were linear detrended and variance was normalized. Spectral analysis toolbox was used to resample data in order to have uniformly distributed data points in time domain, using linear interpolation when needed. Sample size was 1645 and sample frequency was 4.002 samples per kiloyear. We carried out spectral estimation using Thomson's multitaper method with time-halfbandwidth of 1.25 with Thomson's adaptive frequency-dependent weights. For observed data we also calculated 0.95 confidence interval and a noise baseline taking the series as an AR(3) process with Yule-Walker spectral estimation.

\begin{acknowledgements}
 This research was supported by DGAPA-UNAM IN115311 and IN103514 grants, CONACYT-SNI Research Assistant scholarship 3120-8200, as well as by the Asociaci\'on Mexicana de Cultura, A.C.
\end{acknowledgements}

\bibliographystyle{spbasic.bst}
\bibliography{trantor.bib}

\end{document}